\documentclass[twocolumn,prd,superscriptaddress,preprintnumbers,nofootinbib]{revtex4}
\usepackage{graphicx}
\usepackage{epsfig}
\usepackage{lipsum}
\usepackage{enumitem}
\usepackage{bm}
\usepackage{latexsym,amssymb,amsmath,amsfonts,amssymb,txfonts,pxfonts,wasysym,float}
\usepackage{color}
\newcommand{\beq}[1]{\begin{equation}\label{#1}}
\newcommand{\eeq}{\end{equation}}
\newcommand{\bea}[1]{\begin{eqnarray} \label{#1}}
\newcommand{\eea}{\end{eqnarray}}
\newcommand{\ba}{\begin{array}}
\newcommand{\ea}{\end{array}}

\def\be{\begin{equation}}
\def\ee{\end{equation}}
\def\gs{\mathrel{
   \rlap{\raise 0.511ex \hbox{$>$}}{\lower 0.511ex \hbox{$\sim$}}}}
\def\ls{\mathrel{
   \rlap{\raise 0.511ex \hbox{$<$}}{\lower 0.511ex \hbox{$\sim$}}}}

\newcommand{\postscript}[2]{\setlength{\epsfxsize}{#2\hsize}
   \centerline{\epsfbox{#1}}}

\newcommand{\comment}[1]{}

\usepackage[usenames,dvipsnames]{xcolor}
\definecolor{orange}{cmyk}{0,0.5,1,0}
\definecolor{rossoCP3}{cmyk}{0,.88,.77,.40}
\definecolor{graa}{rgb}{0.8,0.8,0.8}
\definecolor{blaa}{rgb}{0.2,0.2,0.6}

\begin{document}

\title{\color{rossoCP3}{Toward a robust inference method for  the likelihood of low-luminosity gamma-ray bursts \\ to be  progenitors of ultrahigh-energy cosmic rays correlating with starburst galaxies}}

\author{Luis A. Anchordoqui}

\affiliation{Department of Physics and Astronomy,  Lehman College, City University of
  New York, NY 10468, USA
}

\affiliation{Department of Physics,
 Graduate Center, City University
  of New York,  NY 10016, USA
}

\affiliation{Department of Astrophysics,
 American Museum of Natural History, NY
 10024, USA
}

\author{Claire Mechmann}

\affiliation{Department of Physics and Astronomy,  Lehman College, City University of
  New York, NY 10468, USA
}

\affiliation{Department of Astrophysics,
 American Museum of Natural History, NY
 10024, USA
}

\author{Jorge F. Soriano}

\affiliation{Department of Physics and Astronomy,  Lehman College, City University of
  New York, NY 10468, USA
}

\affiliation{Department of Physics,
 Graduate Center, City University
  of New York,  NY 10016, USA
}

\begin{abstract}
  \vskip 2mm \noindent Very recently, the Pierre Auger Collaboration
  reported a $4.5\sigma$ correlation between the arrival directions of
  the highest energy cosmic rays and nearby starburst galaxies.  The cosmic rays producing the anisotropy signal have been
  proposed to originate in low-luminosity gamma-ray bursts (llGRBs). On the
  basis of the well-justified assumption that at redshift $z<0.3$ the
  host metallicity is a good indicator of the llGRB production rate, we show
 that the association of llGRBs and the
 starbursts correlating with Auger data is excluded at the 90\% confidence level.
\end{abstract}
\date{December 2019}
\maketitle

\section{General Idea}

By now, it is well-established that galactic-scale outflows of gas
(generally called {\it starburst-driven superwinds}) are ubiquitous in
galaxies in which the global star-formation rate per unit area exceeds
roughly
$10^{-1}~M_\odot~{\rm yr}^{-1} \, {\rm
  kpc}^{-2}$~\cite{Heckman:2000mh}. These flows are complex,
multiphase phenomena powered primarily by massive star winds and by
core collapse supernovae (SNe), which collectively create hot
($T \alt 10^8~{\rm K}$) bubbles of metal-enriched plasma within the
star forming regions. The over-pressured bubbles expand at
high-velocity sweeping up cooler ambient gas and eventually blow out
of the disk into the halo. Starburst superwinds then provide a
commonplace for the formation of collisionless plasma shock waves in
which charged particles can be accelerated by bouncing back and forth
across the shock up to ultrahigh
energies~\cite{Anchordoqui:1999cu}. Experimental data support this
prediction: the Pierre Auger Collaboration reported a $4.5\sigma$
significance correlation between the arrival direction of cosmic rays
with energy above 38~EeV and a model based on a catalog of bright
starburst galaxies~\cite{Aab:2018chp,Aab:2019ogu}. In the best-fit
model, $11^{+5}_{-4}\%$ of the cosmic-ray flux originates from these
objects and undergoes angular diffusion on a scale
$\vartheta \sim {15^{+5}_{-4}}^\circ$. The latter angular spread
derives from a Fisher-Von Mises distribution, the equivalent of a
Gaussian on the sphere, and would correspond to a top-hat scale
$\varphi \sim 1.59 \times \vartheta$.  Of course, readjustment of
superwind-free-parameters are necessary to accomodate Auger
data~\cite{Anchordoqui:2018vji,Anchordoqui:2019mfb,Anchordoqui:2018qom}.

However, it was recently put forward the idea that ultrahigh-energy-cosmic-ray (UHECR) acceleration in low-luminosity gamma-ray bursts (llGRB)
could be the origin of the fraction of Auger events which correlates
with starburst galaxies~\cite{Zhang:2017moz}. In this work we show
that the association of llGRBs with the starbursts generating the
anisotropy signal found in Auger data is disfavored by observation. Before proceeding, we pause to note that whether llGRBs
would satisfy the power requirements to accelerate cosmic rays up to
the highest observed energies may be up for
debate~\cite{Samuelsson:2018fan,Zhang:2018agl,Boncioli:2018lrv}.

The layout of the paper is as follows: in Sec.~\ref{sec:2} we discuss
the sample of llGRBs and starburst galaxies we have selected to study
and conduct the statistical analysis; in Sec.~\ref{sec:3} we draw our conclusions.

\section{Data Analysis}
\label{sec:2}

We begin our study with an overview of the basic properties of the
various GRB populations. A detailed scrutiny of the BATSE catalog led
to our current duration-based classification system for GRBs: short
GRBs (SGRBs) have burst durations of $<2~{\rm s}$, whereas long GRBs
(LGRBs) have burst durations of
$>2~{\rm s}$~\cite{Kouveliotou:1993yx}. GRBs can also be splitted
according to their luminosities into llGRBs
($L_{\rm iso} < 10^{49} ~{\rm erg/ s}$) and high-luminosity GRBs
($L_{\rm iso} > 10^{49} {\rm erg/s}$)~\cite{Liang:2006ci}. Herein, we
also adopt the conventions of~\cite{Levesque:2010qb} to identify
nearby ($z <0.3$) GRBs from those at intermediate redshift
($0.3 < z<1$). We note, however, that recent studies do not show a
strong evidence suggesting that $z < 0.3$ GRBs would be, as a
population, different from the high-redshift one. Nevertheless, although one can
find many high-luminosity GRBs at $z < 0.3$, the llGRBs at $z > 0.3$
cannot be detected due to the sensitivity of the gamma-ray detectors;
see e.g. Fig. 1 in~\cite{Perley:2013kwa}. This last argument further justifies our redshift
selection criterion.

Over the last two decades a consensus formed that LGRBs are a product
of a core-collapse of a massive star~\cite{MacFadyen:1998vz} and that
SGRBs have a different origin. Indeed, observations have proved the
SNe type Ic-BL $\leftrightharpoons$ LGRBs connection beyond any
reasonable
doubt~\cite{Woosley:2006fn,Modjaz:2015cca,Cano:2016ccp}. Type Ic are
core-collapse stripped-envelope SNe, whose progenitor stars have lost
most of the hydrogen and helium in their outer envelopes prior to the
collapse. Some SNe type Ic are found to have very broad lines in their
spectra (type Ic-BL), indicative of very fast ejecta velocities.

Because GRBs are outlying and arise in small galaxies seldom monitored
by high-angular resolution surveys, it has not been and will likely
not be possible in the near future to image the progenitor of a GRB,
thus we are only able to figure out properties of the progenitor star
from its environment. There are several studies that seem to indicate
that GRB formation efficiency drops at high metallicity. For example,
the host galaxies of five nearby LGRBs (980425, 020903, 030329, 031203
and 060218, each of which had a well-documented associated SN) are all
faint and metal-poor compared to the population of local star-forming
galaxies~\cite{Stanek:2006gc}. Moreover, various analyses of GRB host
morphologies suggest a correlation between metallicity and LGRB
occurrence rate; see e.g.~\cite{Kocevski:2009cu,Trenti:2014hka}. In
addition, a systematic comparison of the host galaxies of broad-lined
SNe Ic with and without a detected GRB, indicates that a larger
fraction of super-solar metallicity hosts are found among the SNe
Ic-BL without a GRB~\cite{Japelj:2018sly}.

Models of stellar evolution further reinforce the metallicity bias for
LGRB progenitors.  This is because the well-established correlation
between LGRB and stripped-envelope SNe points to carbon- and
oxygen- rich Wolf-Rayet (WR) stars as the most promising progenitor
candidates~\cite{Woosley:2005gy,Langer:2005hu}.\footnote{WR stars are
  highly luminous massive objects which are at an advanced stage of
  stellar evolution and losing mass at a very high rate.} WR stars emit winds that eject about $10 M_\odot$ of
material per million years at speeds of up to $3,000~{\rm km/s}$,
resulting in the characteristic broad emission lines in the spectra of
these stars (normal stars have narrow emission lines). It is thought that these powerful winds are driven
by intense radiation pressure on spectral lines, yielding a dependence of
the wind-driven mass loss rate on surface
metallicity~\cite{Kudritzki:2002ni,Vink}. Thereupon, the
 surface
rotation velocities of WR stars are expected to decrease at higher stellar
metallicites because of the higher mass loss
rate~\cite{Kudritzki:2000kg}. For WR stars, the metallicities
characterizing their host environments can be adopted as the natal
metallicities of the stars themselves. This entails that the higher wind-driven mass loss rates in metal-rich environments
would remove from the massive WR stars too much angular momentum, inhibiting
them from rotating rapidly enough to produce a
LGRB~\cite{Woosley:2005gy,Yoon:2012aq}. All in all, the data seem to
indicate that LGRBs should be confined to low-metallicity environments.

Though {\it a priori} there is no reason to assume that LGRBs and
llGRBs are related, the similarity of their associated SNe implies
that llGRBs and LGRBs have similar progenitors and similar inner
explosion mechanism~\cite{Nakar:2015tma}.  In light of the preceding
discussion, it seems reasonable to assume that the metallicity of the
host environment would also be a good discriminator of llGRB
progenitors. In what follows we compare the host metallicity of nearby
llGRBs with that of the starbursts dominating the signal in Auger data.

Before we can conduct the statistical analysis, we need to define our
samples. The Auger anisotropy search included a sample of 23 starburst
galaxies with a flux larger than 0.3~Jy selected out of the 63 objects within 250 Mpc search for
$\gamma$-ray emission by the Fermi-LAT Collaboration~\cite{Ackermann:2012vca}. This selection was updated in~\cite{Aab:2019ogu} 
with the addition of the Circinus Galaxy and sources selected from the
HEASARC Radio Master Catalog.\footnote{\tt https://heasarc.nasa.gov/W3Browse/master-catalog/radio.html} The number of starbursts selected this way
is 32. Here we consider 10 of these galaxies (including all sources
dominating the Auger anisotropy signal) for which the average
metallicity has been determined. It is important to note that some
galaxies in the starburst sample have a double starburst/AGN nature
(e.g., Circinus, NGC 4945, NGC 1068). Given that so far, despite
efforts, no GRB host galaxy has been found to host an AGN, it may not
come as surprising if the two samples are not drawn from the same
underlying probability distribution. We consider all llGRB detected at
$z<0.3$. The metallicities of llGRB hosts are given in
Table~\ref{tabla:1} and the metallicities of the starbursts are given in
Table~\ref{tabla:2}.  Following~\cite{Moreno-Raya:2016rlw}, we have
taken $\log(Z/Z_\odot) = \log ({\rm O/H}) - \log({\rm O/H})_\odot$,
with $12 + \log ({\rm O/H})_\odot = 8.69$ and $Z_\odot = 0.019$ being
the solar values~\cite{Asplund:2009fu}.\footnote{We note that the
  precision of our phenomenological study is insensitive to any
  plausible change of the solar metallicity, e.g.,
  $12 + \log ({\rm O/H})_\odot = 8.66$~\cite{Asplund:2003ws}.}

\begin{table}
  \caption{Properties of nearby llGRBs. \label{tabla:1}}
\begin{tabular}{lcccc}
\hline
\hline
GRB ID & $\log[L_{\rm iso}/({\rm erg/s})]$  & Redshift & $12+\log({\rm O/H})$ & References \\
\hline
980425 & $46.67$ & 0.008 & $8.3$ &  \cite{Stanek:2006gc,Levesque:2010qb,Virgili:2008gp}  \\
020903 & $48.92$ & 0.251 &  $8.0$ &  \cite{Levesque:2010qb,Virgili:2008gp,Thone:2019epb}  \\
031203 & $48.55$ & 0.105 & $8.1$ &  \cite{Levesque:2010qb,Virgili:2008gp,Thone:2019epb}\\
  051109B & 48.22   & $0.080$ & $\cdots$
& \cite{Pescalli:2014qja} \\
060218 & $46.78$ & 0.033 & $8.1$ &  \cite{Levesque:2010qb,Virgili:2008gp, Thone:2019epb} \\
060505  & 48.85 & 0.089 &  $8.4$ & \cite{Pescalli:2014qja,Thone:2007ee} \\
080517 & $48.52$ & 0.089 & $8.6$ & \cite{Li,Niino:2016ogw,Stanway:2014via,Chrimes:2018ptj} \\
100316D & $47.75$ & 0.059 & $8.2$ &  \cite{Levesque:2011ph,Dereli:2017,Thone:2019epb}\\
111005A &          $46.78$      &    0.013          & $8.6$ & \cite{Li,Tanga:2017obk,Michalowski:2016hzq}\\
171205A & $47.50$ & 0.037 & $8.4$ &
                                    \cite{Izzo:2019akc}\footnote{An
                                    estimate of the GRB 171205A host
                                    metallicity is given in the
                                    journal (but not in the arXiv) version of~\cite{Izzo:2019akc}.}\\
\hline
\hline
\end{tabular}
\end{table}
\begin{table}
 \caption{Properties of nearby starburst galaxies. \label{tabla:2}} 
\begin{tabular}{lccc}
\hline
\hline
Starburst ID~~ & ~~Distance (Mpc)~~ & ~~12+ $\log({\rm O/H})$~~ & ~References~ \\
\hline
  NGC 253 & $\phantom{1}2.7$ & 8.7 &  \cite{Aab:2018chp,Zaritsky:1994ht,Pilyugin:2004sc}\\
  M82 & $\phantom{1}3.6$ & 8.8 &
                                 \cite{Aab:2018chp,Moreno-Raya:2016rlw}\footnote{See,
  in particular, Table~A.1 of~\cite{Moreno-Raya:2016rlw}.}\\ 
  NGC 4945 & $\phantom{1}4.0$ &  8.5  &  \cite{Aab:2018chp,Stanghellini}\\
  M83   & $\phantom{1}4.0$ & 8.8 & \cite{Aab:2018chp,Zaritsky:1994ht,Pilyugin:2004sc} \\
IC 342 & $\phantom{1}4.0$ & 8.8 &  \cite{Aab:2018chp,Pilyugin:2004sc,Hung}\\
Circinus & $\phantom{1}4.0$ & 8.4 & \cite{Oliva:1998bf}\footnote{See,
  in particular, Table~1 of~\cite{Oliva:1998bf}.} \\
  NGC 6946 & $\phantom{1}5.9$ & 8.8 & \cite{Aab:2018chp,Zaritsky:1994ht,Pilyugin:2004sc} \\
  M51 & 10.3 & 8.8 &  \cite{Aab:2018chp,Zaritsky:1994ht,Pilyugin:2004sc} \\
  NGC 891 & 11.0 & 8.7 &  \cite{Aab:2018chp,Galliano:2007nw,Otte:2001np} \\
NGC 1068 & 17.9 & 8.8 & \cite{Aab:2018chp,Zaritsky:1994ht,Pilyugin:2004sc} \\
  \hline
  \hline
\end{tabular}
\end{table}

Before proceeding, some technical remarks are in oder to clarify our
metallicity selection criteria. Molecular gas in starbursts exists
under conditions very different from those found in most normal
galaxies. Observations of starbursts suggest widespread gas volume and
column densities much higher than those typical of normal
disks~\cite{Jackson:1995,Paglione:1997}. The nebular oxygen abundance
is the canonical choice of metallicity indicator for studies of the
interstellar medium since oxygen is the most abundant metal, only
weakly depleted, and exhibits very strong nebular emission lines in
the optical wavelength range~\cite{Tremonti:2004et}. Extensive analyses have been carried
out to calibrate metallicity studies by using only strong emission lines. One of the most frequently used metallicity
 diagnostics is the parameter 
\begin{equation}
R_{23} = \log_{10} \left\{([{\rm O}\textsc{ii}] \lambda3727 +
  [{\rm O}\textsc{iii}] \lambda \lambda4959, 5007)/{\rm H}\beta \right\}, 
\end{equation}
defined as the ratio of the flux in the strong optical oxygen lines to that of
 H$\lambda4861$~\cite{Pagel:1979vx}; notation conventions are those
 in~\cite{Tendulkar:2017vuq}. However, a well-known problem of this metallicity
 diagnostic is that the $R_{23}$ vs. $12 + \log({\rm O/H})$ relation is double-valued, and so additional information is required to break
 this degeneracy. Several methods have been developed to remove the
 $R_{23}$ degeneracy exploiting the [N{\sc ii}], [S{\sc ii}], and H$\alpha$ lines;  e.g.,
 \begin{eqnarray}
 N2 & = & \log_{10} \left\{[{\rm N} \textsc{ii}] \lambda6584/{\rm H} \alpha \right\} \,,
          \nonumber \\
   O3N2 & = &\log_{10} \left\{([{\rm O}\textsc{iii}]\lambda5007/[{\rm N}
              \textsc{ii}] \lambda6584) \times ({\rm H}\alpha/{\rm H}\beta) \right\} \,,
              \nonumber \\
    y &= & \log_{10} \left\{[{\rm N}\textsc{ii}]
           \lambda6584/[{\rm S}\textsc{ii} \lambda\lambda 6717,6731 \right\}
           \nonumber \\
    & + & 0.264 \log_{10} \left\{[{\rm N} \textsc{ii}]
          \lambda6584/{\rm H}\alpha
          \right\} \,,
          \end{eqnarray}
          proposed in~\cite{Kewley:2002cv}, \cite{Pettini:2004gk},
          and~\cite{Dopita:2016rqe}.  Although an absolute calibration
          for metallicities obtained through the strong-line methods
          remains uncertain~\cite{Kewley:2008mx}, we may still use the
          strong-line ratios to study the trend in metallicities
          between the llGRB hosts and starburst galaxies in our
          sample. Indeed, the absolute metallicity scale varies up to
          $\Delta[\log({\rm O/H})] = 0.7$, depending on the
          calibration used, and the change in shape is substantial. It
          is critical then to use the same metallicity calibration
          when comparing different metallicity relations. Herein we
          adopt the $O3N2$ diagnostic with normalization as given
          in~\cite{Pettini:2004gk},
          \begin{equation}
 12 + \log ({\rm O/H}) = 8.73 - 0.32 \times O3N2\,,
\end{equation}
         and use the metallicity conversions
          given in~\cite{Kewley:2008mx}, which allow metallicities
          that have been derived using different strong-line
          calibrations to be converted to the same base
          calibration. In Tables~\ref{tabla:1} and \ref{tabla:2} we provide the best-fit values of the metallicities after
          conversion to the same base calibration.
Following~\cite{Galliano:2007nw}, an uncertainty of 0.1 dex in the O/H number abundance
accounts for the typical dispersion between independent
measurements. To remain conservative, in our calculations we adopt the
upper and lower end of the $1\sigma$ metallicity range to characterize the llGRB
and starburst samples, respectively. Concerning GRB 051109B,
it has been tentatively associated with a star-forming region in a
          spiral galaxy which lacks of any strong emission features~\cite{Perley}. Therefore, we do not include this event in our statistical analysis.

\begin{figure}[tb] 
  \postscript{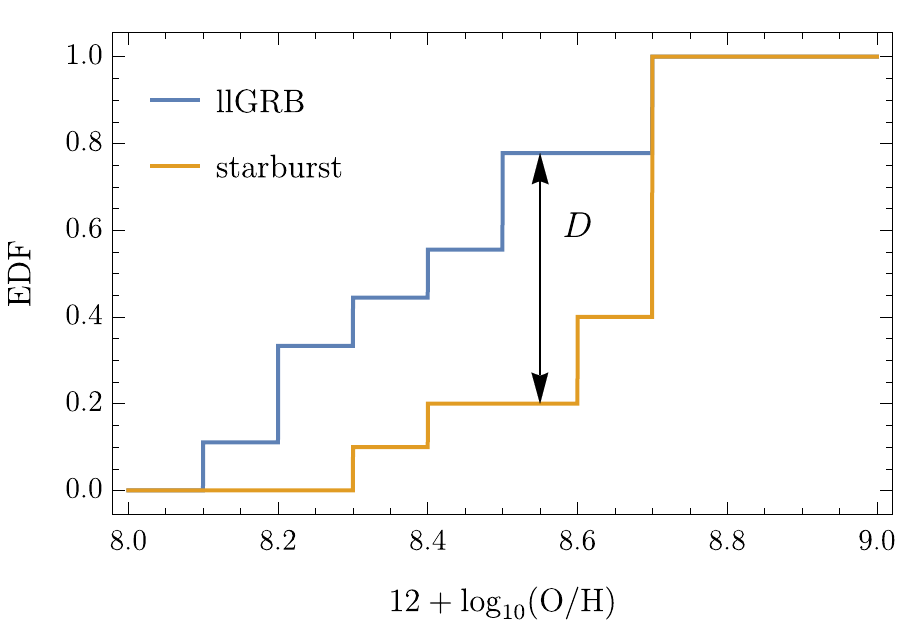}{0.9} 
\caption{Vertical displacement between sample distribution functions. \label{figura}}
\end{figure}

Next, we adopt the Kolmogorov-Smirnov (two-sample) test to check whether the
two data sets of metallicity are both drawn from the same underlying
probability distribution, but without assuming any specific model for
that distribution~\cite{Kolmogorov,Smirnov}. The calculations that are involved in application
of the Kolmogorov-Smirnov test are quite simple.  We begin by stating
the null hypotheis ${\cal H}_0$: if $f_m(x)$ and $g_m(x)$ are samples
of two underlying probability density functions $f(x)$ and $g (x)$,
then
\begin{equation}
  {\cal H}_0: ~f(x) = g(x) \,, \forall x \, .
\end{equation}
The alternate hypothesis is that $f(x) \neq g(x)$. Now, given any
sample from an unspecified population, a natural estimate of the
unknown cumulative distribution function of the population is the {\it
  empirical (or sample) distribution function} (EDF) of the sample, defined, at any real number $x$, as the proportion of sample observations which do not exceed $x$. For a sample of size $m$, the empirical distribution function will be denoted by $F_m(x)$ and may be defined in terms of the order statistics $X_{(1)}\leq X_{(2)} \cdots \leq X_{(m)}$ by
\begin{equation}
F_m(x) = \left\{ \begin{array}{ll}
0 & ~~{\rm if} \ x < X_{(1)} \\
j/m & ~~{\rm if} \  X_{(1)} \leq x \leq X_{(j+1)}, \ 1 \leq j \leq m \\
1 & ~~{\rm if} \  x \geq X_{(m)}
\end{array}
\right. \, ,
\end{equation}
i.e., $F_m$ is the {\it staircase function}.

To form the test statistics $D$ from the sample distribution functions
$F_m(x)$ and $G_n (x)$ we compute their maximum absolute difference
over all the values of $x$,
\begin{equation}
  D = \max_{x} \ \left|F_m(x) - G_n(x) \right| \, .
\end{equation}
Graphically, we may interpret this as the maximum vertical
displacement between the two sample distribution functions as
indicated in Fig.~\ref{figura}.

Testing of the null hypothesis proceeds by comparison of $D$ against
critical values $D_\alpha$ which are functions of the confidence level $\alpha$
and the sizes of the samples $m,n$~\cite{Conover}. We may reject the null hypothesis ${\cal H}_0$ at the
$(1-\alpha)$ confidence level
if $D > D_\alpha$.  For the case at hand, $m= 9$ and $n=10$, the  upper critical
 value of the 90\%
confidence level interval is $D_{0.1} = 5/9$~\cite{Rohlf}.  Since the
 maximum difference between the EDFs shown in Fig.~\ref{figura} is
 $D = 26/45$, we infer that the null-hypothesis (the two metallicity samples
 belong to the same distribution) is excluded at the 90\% confidence
 level. Therefore, on the basis of the well-justified assumption that at redshift \mbox{$z<0.3$} the
  host metallicity is a good indicator of the llGRB production rate, we can conclude
 that the association of llGRBs and the starbursts correlating with
 Auger data is disfavored by observation.

\section{Conclusion}
\label{sec:3}

We have used the metallicity of the llGRB host galaxies as a proxy to
investigate whether llGRBs can be the sources of the highest energy
cosmic rays whose arrival directions correlate with the celestial
positions of nearby starburst galaxies. We have shown that the
association of llGRBs and the starbursts correlating with Auger data
is excluded at the 90\% confidence level. We end with two
observations:
 \begin{itemize}[noitemsep,topsep=0pt]
 \item The first one builds upon the estimates in~\cite{Pfeffer:2015idq} and contributes to
   the debate on the source power requirements. The Telescope Array 
   Collaboration has reported an excess of UHECR events over expectations
   from a random distribution in a circle of $20^\circ$ near M82~\cite{Abbasi:2014lda}. The hotspot energy flux
   in UHECRs with energies $E > E_0 = 5.7 \times 10^{10}~{\rm GeV}$ is
   estimated to be
   \begin{eqnarray}
     F_{\rm hs} & = & \Omega_{20^\circ} \int_{\rm E_0}^\infty E  \
                      J_{\rm hs} (E) \ dE \nonumber \\
& \simeq &
     1.7 \times 10^{-8} \xi_{1.7}~({\rm GeV  \ cm^2 \ s})^{-1}    \,,
\end{eqnarray}   
where $\Omega_{20^\circ} \simeq 0.38$ is the hot-spot solid
angle and $\xi_{1.7}$ parametrizes the uncertainty in the energy
dependence of
 the specific (number) intensity in the hotspot
$J_{\rm hs} (E)$~\cite{Pfeffer:2015idq}.  The rms deflection angle for an UHECR of charge $Ze$ is found to be
     \begin{equation}
       \delta_{\rm rms} \approx 3.6^\circ \ Z \ E_{11}^{-1} \ r_{\rm
         kpc}^{1/2} \ \lambda_{\rm kpc}^{1/2} \ B_{\mu{\rm G, rms}} \,,
     \end{equation}
 where $B_{\mu{\rm G,rms}}$ is the rms strength of the magnetic field
   in $\mu$G, $E_{11}$ is the UHECR energy in units of $10^{11}~{\rm
     GeV}$, $r_{\rm kpc}$ is the distance over which the
   magnetic fields act in kpc, and $\lambda_{\rm kpc}$ is the magnetic-field
   coherence length also in units of kpc~\cite{He:2014mqa}. The scattering
   in the magnetic field
   also gives a time spread~\cite{Waxman:1995vg,Farrar:1999bw}, which is given by
   \begin{eqnarray}
  \tau & \simeq & 4.1 \ \left(\frac{r_{\rm kpc} B_{\mu{\rm
                  G}}}{E_{11}}\right)^2 \lambda_{\rm kpc} Z^2~{\rm yr} 
                  \nonumber \\
       & \simeq & 4.1 \ \left(\frac{\delta_{\rm rms}}{3.6^\circ}\right)^2 r_{\rm kpc}~{\rm yr} \, .
   \end{eqnarray}
Because all the UHECR scattering occurs inside the Galaxy, we have
$r_{\rm kpc} \sim 10$ and $\delta_{\rm rms} \sim 20^\circ$, yielding 
   a dispersion of $\tau \sim 10^3~{\rm yr}$ in the UHECR arrival
   times.  For a source at a distance $D$, the required isotropic
   equivalent luminosity is
   $L_{\rm iso} = 4 \pi D^2 F_{\rm hs}$ and so the
   isotropic-equivalent energy implied is $E_{\rm iso} > 10^{51} \ \xi_{1.7}~{\rm
     erg}$. It is noteworthy that
   llGRBs struggle to meet this constraint as they all have $E_{\rm iso} <
   10^{50}~{\rm erg}$~\cite{Cano:2016ccp,Stanek:2006gc}. This is also the case for SNe with 
   relativistic outflows but without GRB counterparts, for which the
   observed 
   isotropic-equivalent energy is on the order of $10^{49}~{\rm
     erg}$~ \cite{Soderberg:2009ps,Chakraborti:2014dha,Margutti:2014gha}.  
 \item We now comment on the possibility that the main hypothesis of
   our analysis is false; namely, that the host metallicity {\it is
     not} a good indicator of the llGRB production rate (see
     e.g.~\cite{Kelly:2014fra,Perley:2016bke,Palmerio:2019nvq}  for
   a discussion on other considerations that could affect the LGRB production
               efficiency). In this direction, it is {\it natural} to envision the most straightforward scenario, in which
the llGRB rate is independent of all factors other than the overall
rate of star-formation itself. This would imply that  a fixed
fraction of all newly-formed stars could explode as llGRBs without
perception to any of the chemical, physical, or other properties of
the galaxy in which those stars formed. From the observational
viewpoint, this entails that llGRBs should stochastically sample the
locations of cosmic star-formation throughout the volume of the
Universe in which they can be observed. The probability that any given
galaxy will host a llGRB during some period of time would then be proportional to
its star-formation rate. Now, given the ubiquity of llGRBs in this
simplistic scenario we
   can ask ourselves why the correlation of UHECRs with starburst
   galaxies would be explained by the presence of this {\it common}
   phenomenon. Rather there must be some other inherently unique
   feature(s) of starburst galaxies to account for this
   correlation. Starburst galaxies represent about 1\% of the fraction
   of galaxies containing star forming galaxies~\cite{Bergvall}, and
   the probability of SN explosions is about an order of
   magnitude larger in starbursts than in normal galaxies, e.g., the
   SN rate for M82 is about $0.2-0.3~{\rm yr}^{-1}$~\cite{Ulvestad:1995yt} whereas for the Milky
   Way is $\sim 3.5 \pm 1.5~{\rm century}^{-1}$~\cite{Dragicevich}. Note that these two effects tend
   to compensate each other, and so if the anisotropy signal reported
   by the Auger Collaboration originates in llGRBs (within this particular
   underlying scenario), then when studying the
   correlation of UHECRs with the nearby matter distribution the
   statistical significance must increase.  However, when all sources
   beyond 1~Mpc (i.e. effectively taking out the Local Group) from
   the 2MRS catalog are included as part of the anisotropic signal in
   the analysis of~\cite{Aab:2019ogu} the significance level reduces
   from $4.5\sigma$
   to $3.8\sigma$. Altogether, the data yielding the anisotropy signal seem to favor a production
   mechanism of UHECRs above 38~EeV which
   is exclusive to starbursts, like Fermi-shock acceleration in
   starburst superwinds.
\end{itemize}

  \acknowledgments{We thank Michael Unger for useful critique. We also thank our colleagues of the Pierre Auger
    and POEMMA collaborations for some valuable discussion. This work
    has been supported by the by the U.S. National Science Foundation
    (NSF Grant PHY-1620661) and the National Aeronautics and Space
    Administration (NASA Grant 80NSSC18K0464).  Any opinions,
    findings, and conclusions or recommendations expressed in this
    material are those of the authors and do not necessarily reflect
    the views of the NSF or NASA.}

\end{document}